\documentclass[10pt,letterpaper]{article}
\usepackage[top=0.85in,left=1.75in,footskip=0.75in,marginparwidth=2in]{geometry}

\usepackage[utf8]{inputenc}

\usepackage{cite}
\usepackage{longtable}
\usepackage{threeparttable,tablefootnote}
\usepackage{amsthm}
\usepackage[english]{babel}

\theoremstyle{definition}

\usepackage{nameref,amsmath,amsfonts}

\usepackage[right]{lineno}
\usepackage[colorlinks=true,citecolor=blue,linkcolor=blue]{hyperref}%

\usepackage{microtype}
\DisableLigatures[f]{encoding = *, family = * }

\raggedright
\usepackage{changepage}
\usepackage{threeparttable, tablefootnote}

\usepackage[aboveskip=1pt,labelfont=bf,labelsep=period,singlelinecheck=off]{caption}

\makeatletter
\renewcommand{\@biblabel}[1]{\quad#1.}
\makeatother

\usepackage{lastpage,fancyhdr,graphicx}
\usepackage{epstopdf}
\fancyheadoffset[L]{2.25in}
\fancyfootoffset[L]{2.25in}

\usepackage{color}

\definecolor{Gray}{gray}{.25}

\usepackage{graphicx}

\usepackage{sidecap}

\usepackage{wrapfig}
\usepackage[pscoord]{eso-pic}
\usepackage[fulladjust]{marginnote}
\reversemarginpar

\begin{document}
\vspace*{0.35in}

\begin{flushleft}
{\Large
\textbf\newline{Semi-Supervised Record Linkage  for  Construction of Large-Scale Sociocentric Networks in Resource-limited Settings: An application to the SEARCH Study in Rural Uganda and Kenya.}
}
\newline
\\
Yiqun Chen\textsuperscript{1}, Wenjing Zheng\textsuperscript{1}, Lillian B. Brown\textsuperscript{2}, Gabriel Chamie\textsuperscript{2}, Dalsone Kwarisiima\textsuperscript{3}, Jane Kabami\textsuperscript{3}, Tamara D. Clark\textsuperscript{2}, Norton Sang\textsuperscript{4}, James Ayieko\textsuperscript{4}, Edwin D. Charlebois \textsuperscript{2,6}, Vivek Jain \textsuperscript{2,6}, Laura Balzer\textsuperscript{5}, Moses R Kamya \textsuperscript{7}, Diane Havlir\textsuperscript{2}, Maya Petersen\textsuperscript{1,*}, and the SEARCH Collaboration 
\\
\bigskip
\bf{1} Division of Biostatistics, University of California, Berkeley,  CA USA
\\
\bf{2} Division of HIV, ID, \& Global Medicine, University of California, San Francisco, CA USA
\\
\bf{3} Infectious Disease Research Collaboration, Kampala, Uganda
\\
\bf{4}  Kenya Medical Research Institute, Nairobi, Kenya
\\
\bf{5} Department of Biostatistics and Epidemiology, University of Massachusetts, Amherst, MA. 
\\
\bf{6}  Center for AIDS Prevention Studies, University of California, San Francisco, CA USA
\\
\bf{7} Makarere University, Kampala, Uganda \& Infectious Disease Research Collaboration, Kampala, Uganda\\
\bigskip
* mayaliv@berkeley.edu
\end{flushleft}
\newpage
%
\section*{Abstract}
This paper presents a novel semi-supervised algorithmic approach to creating large scale sociocentric networks in rural East Africa.
We describe the construction of 32 large-scale sociocentric social networks in rural Sub-Saharan Africa. Networks were constructed by applying a semi-supervised record-linkage algorithm  to data from census-enumerated residents of the 32 communities included in the SEARCH study (NCT01864603), a community-cluster randomized HIV prevention trial in Uganda and Kenya. Contacts were solicited using a five question name generator in the domains of emotional support, food sharing, free time, health issues and money issues. The fully constructed networks include $170,028$ nodes and $362,965$ edges aggregated across communities (ranging from 4449 to 6829 nodes and from 2349 to 31,779 edges per community). Our algorithm matched on average $30\%$ of named contacts in Kenyan communities and $50\%$ of named contacts in Ugandan communities to residents named in census enumeration. Assortative mixing measures for eight different covariates reveal that residents in the network have a very strong tendency to associate with others who are similar to them in age, sex, and especially village. The networks in the SEARCH Study will provide a platform for improved understanding of health outcomes in rural East Africa. The network construction algorithm we present may facilitate future social network research in resource-limited settings.

\section*{Highlights}
\begin{itemize}
\item 
We describe a systematic network construction and analysis pipeline from raw census data that might be applicable in resource-constrained settings, in which an absence of an unambiguous unique identifiers for linking enumerated participants to named contacts.
\item
32 large sociocentric networks of size around 5,000 each from  communities in Uganda and Kenya are presented. To our knowledge, this is the largest set of sociocentric networks linked to health data to be constructed in Sub-Saharan Africa.
\item
In a proof-of-concept analysis of the resulting networks, an individual was found to be much more likely to associate with others who are similar to him or her in age, village, and sex.

\end{itemize}

\section*{Keywords}
Sociocentric network, network construction, record linkage, assortative mixing

\section{Introduction}

 Social networks contribute extensively to understanding human behavior and relations, particularly in public health settings. A growing body of literature in epidemiology, in particular, investigates behavioral determinants of health-related outcomes and the role of social networks in shaping individual health conditions \cite{Tsai_Perkins2018-is,Christakis2008-rc,Christakis2007-ir,Danon2011-jl,Salathe2010-bp,Riley2007-ur,Klovdahl1994-zz,Eubank2004-bb,Keeling2005-tz,Kramer2014-my,Christakis2013-tk,Kossinets2006-rv,Choice1998-ny}. Prior work has measured specific social connections representing participants' personal ties, analyzed how those ties can be used as proxy measures for transmission networks for infectious diseases or support networks for non-communicable diseases\cite{Christakis2008-rc,Christakis2007-ir,Christakis2013-tk,Christley2005-yy,Newman2002-rq,Kretzschmar1996-vc,Fowler2008-qb,Fowler2008-rm}, and explored the use of social networks to improve intervention effectiveness, stratify risk, and inform precision public health policies\cite{Eubank2004-bb,Lipsitch2003-wo,Klovdahl1985-ta,Read2008-nh,Fowler2008-qb,Loucks2006-dj,Cohen-Cole2008-gw}.
 
Much of the social network construction and analysis reported in the epidemiologic and health literature have been conducted in resource-rich settings. In sub-Saharan Africa, in particular,  literature describing the construction and analysis of networks is limited \cite{Perkins2015-dm,Tsai_Perkins2018-is}. 
Moreover, much of the existing literature  focuses on ego-centric networks, constructed by interviewing participants and their named contacts using methods such as respondent-driven sampling or linkage tracing
\cite{Thompson2000-fk,Gile2015-gm,Lin2012-jn}. In contrast, sociocentric networks aim to map entire collections of social relations for surveyed communities, and therefore offer opportunities to answer a richer set of research questions using global network characteristics, such as overall network densities, centralities, and local topologies; such properties have been found in both simulations and applied analyses to be important determinants of health behavior\cite{Christley2005-yy,Perkins2015-dm,Chami2014-dj,Amirkhanian2014-xy,Kossinets2006-rv}. 

To date, sociocentric networks that have been constructed in  resource-limited settings have generally consisted of  a single community of small to moderate size \cite{Shah2017-tj,Chami2014-dj,Rothenberg2009-el,Kelly2014-vw}. Networks in the present HIV literature, in particular,  have been commonly based on networks of approximately 50-100 individuals \cite{Lin2012-jn,Chami2014-dj,Rothenberg2009-el,Morris2004-xi,Read2008-jl,Latkin2015-te}. Larger sociocentric networks in this setting include the Likoma network study conducted in Likoma Island,  with a total size of 923 nodes (participants) and 2040 edges consisting of sexual relationships\cite{Likoma}, and more recently constructed sociocentric networks in rural Uganda with 1669 nodes \cite{Takada2019-wr} and Tanzania with 923 nodes \cite{Yamanis2017-kw,Mulawa2018-vd}.

One of the factors that may contribute to the rarity of large-scale sociocentric networks in resource-limited settings is a lack of explicit algorithms for network construction. While research on methods for  analyzing social network data is extensive, much less work has been explicitly dedicated to the construction of sociocentric networks, especially for large communities where real-time linkage of participants and named contacts is not feasible. The current paper aims to contribute to the literature on algorithms for large-scale sociocentric network construction in resource-limited settings by describing a record linkage-based  semi-supervised algorithm that enables network construction from raw census data consisting of participants and their named contacts. 

The algorithm is applied to baseline social network data from the SEARCH (Sustainable East Africa Research in Community Health) Study (NCT01864603), a cluster-randomized HIV-prevention study that measured HIV infection, additional  health outcomes, and health-related behaviors longitudinally in approximately 320,000 individuals. Data from SEARCH are used to construct 32 distinct sociocentric social networks in three different geographical regions in rural Kenya and Uganda, with approximately 5,000 adult participants in each network. As in many studies, the absence of unique identifiers for named contacts in the study creates a substantial linkage challenge. We calibrate the performance of our algorithms by computing the percent of all named contacts matched, a metric rarely reported in the literature \cite{Rothenberg2009-el}.  We also provide metrics for evaluating linkage performance, describe characteristics of the resulting networks, and report a proof-of-concept assortative mixing analysis, which not only provides information about the structure of a network, but also informs inference on the 
and robustness properties of the network \cite{Lemieux-Mellouki2016-lz,Bollen2011-eo}.
%

\section{Methods}

\subsection{Data collection:}
The  Sustainable East Africa Research in Community Health (SEARCH) Study is a community cluster-randomized trial (NCT01864683) in 12  communities in rural Kenya and 20 communities in rural Uganda (10 in the southwestern region and 10 in the eastern region). Each community contains approximately 10,000 residents, of whom about 50\% are adults ($\geq15$ years of age). A door-to-door household census was conducted at study baseline  (June 2013 – June 2014), and involved collection of demographic information from all residents including age, sex, marital status, education level, income level, and occupation. Census enumeration was followed by HIV and multi-disease testing using a hybrid model that combined multi-disease community health campaigns with home-based testing for non-attendees, and which reached 89\% of the enumerated population \cite{Chamie2016-sd}. During hybrid testing, adult (aged $\geq 15$ years) residents were asked to name up to 6 contacts in each of 5 different social domains (food issues, health issues, monetary matters, interaction during free time, and emotional support), as well as the age and village of these contacts, adapted from earlier work by \cite{Tsai_Perkins2018-is}. For instance, regarding money domain, the question is phrased as ``Over the last 12 months, with whom have you usually discussed any kind of money matters?". A complete set of name generators is included in the Appendix. Contact names were collected by field staff on tablet computers. All participants provided verbal informed consent in their preferred language with fingerprint confirmation of agreement. The Makerere University School of Medicine Research and Ethics Committee (Uganda), the Ugandan National Council on Science and Technology (Uganda), the KEMRI Scientific and Ethics Review Unit (Kenya), and the UCSF Human Research Protection Program \& IRB (USA) approved the consent procedures and the study.

\subsection{Network Construction:}
Using these raw data, sociocentric networks for each community were constructed using the following steps: pre-processing, two-step blocked matching, optimal weight determination, and post-processing (Figure 1), each described in the following sections in detail.  Computing code is publicly available (\url{https://github.com/yiqunchen/SEARCH_code}). While only adult residents were able to name contacts, matching was performed for all enumerated residents, allowing children named as contacts to be matched.

\begin{figure}[ht!]
\centering
\includegraphics[scale=0.4]{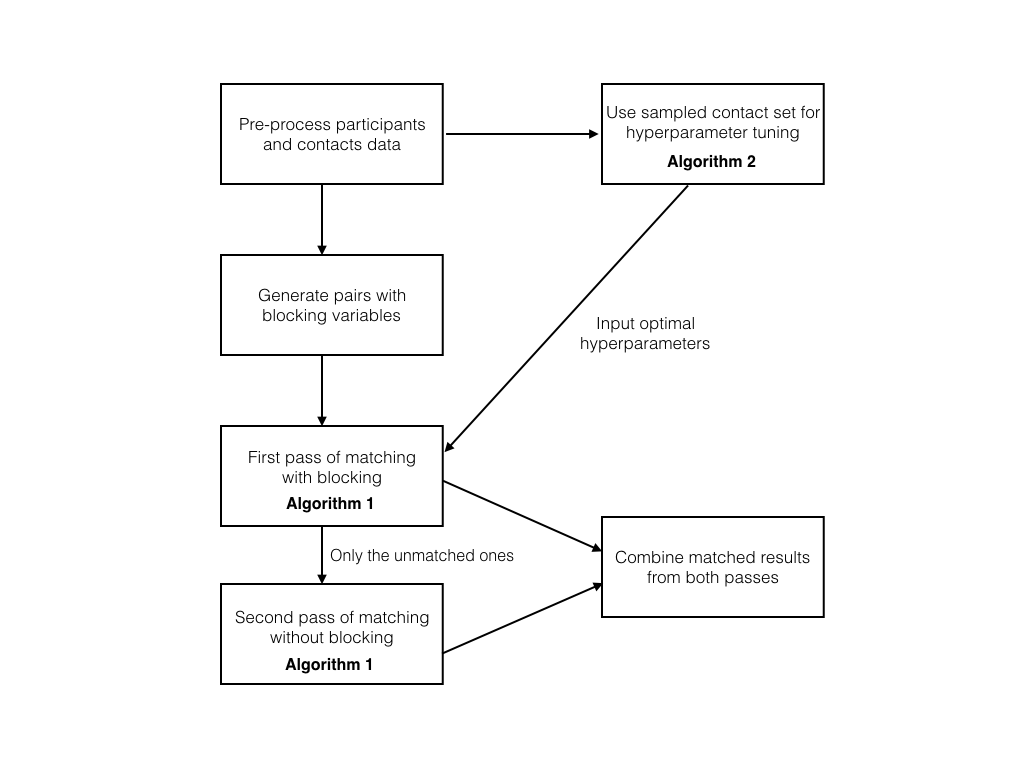}
\caption{Flowchart of network construction}
\end{figure}
\subsubsection{Pre-processing of census data:}

Building on record linkage algorithms reported in the literature 
\cite{Randall2013-mb,Christen2012-bm}, we applied the following techniques in our pre-processing pipeline.

\begin{enumerate}
\item Name Standardization:
\begin{enumerate}
\item 
 Remove extra spaces, remove punctuation within name fields, and combine names  to a unique string.

\item
Nickname Lookups: In Kenyan communities, common nicknames were identified and pre-processed. For example,  ``Min", ``Wuon", ``Nyar" and ``Nya",  stand for, respectively,``Mother of", ``father of",  ``daughter of", and``young one". For instance, someone referred to as  Nyar John is ``daughter of John".
\item
 Honorific and suffix lookups: Similar to nicknames, we extract honorifics and suffix information using a lookup table (including Dr./Mr./Jr. and etc.).
\item
Breakdown of name fields: Names of participants and contacts are then broken down into up to 4 individual components (excluding nicknames, honorifics add suffices).
\item
Permutation of name fields: The name fields are then permuted to maximize chances of match. For instance, Jason Max Nissima would appear in all six of its possible permutations, as $3! = 6$

\end{enumerate}
\item Sex imputation: 
As participants were not asked about the sex information of their contacts, all records of contacts have a missing sex value. Sex of contacts were imputed using a lookup table equating common first names with sex. 
\item Address Standardizations:
Unique villages listed by participants and contacts were compared and fixes to villages of contacts were made according to the specific data. Common fixes, for instance, would include numeral standardization such as changing “Nsiika II” into “Nsiika 2” and common typos such as “Nsika” into “Nsiika”.

\end{enumerate}

\subsubsection{Pair linking and matching:}
The cleaned and standardized files were then matched using a semi-supervised learning algorithm with the following steps: parameter tuning, linking and matching with blocking, 
linking and matching without blocking, and finally, post-processing.
 
The matching process is broken into three algorithms. Algorithm 1 serves as the template for finding a match for given hyperparameters; algorithm 2 provides a guide on hyperparameter tuning and finally algorithm 3 is the meta-algorithm summarizing the streamline in the matching process, which calls both of the previous algorithms.
\begin{itemize}
\item 
\textbf{Algorithm 1: General Linking and Matching strategy}

\textit{Input: }Cleaned participant and contact data sets, a weight vector $w$ for fields we want to match on (in this application, a vector in 7 dimensional simplex) and a quantile $q$ for fitting Pareto distribution.

\textit{Output: }Matched pairs

\begin{enumerate}
\item
First linking stage: our algorithm starts with a blocking procedure on first name, middle name, last name, village name and sex, i.e., only participants and contacts with exact the same entry on at least one of the five fields will be linked and considered for the following steps.
\item
We then compute the similarities between seven different fields: first name, middle name, last name, age, village, sex and honorifics and/or suffixes. The similarity for age is defined to be $1-\frac{|age_2-age_1|}{100}$, while the other fields are treated as two strings $s_1,s_2$, and Jaro-Winkler (JW) similarity is used. JW similarity, denoted by $S_{JW}$, is defined based on Jaro distance $d_{Jaro}$:
%
%
%
%
 \begin{align}
d_{Jaro}(s_1,s_2) = \begin{cases} 
      0 & \text{if } m = 0\\
      \frac{1}{3}(\frac{m}{|s_1|}+\frac{m}{|s_1|}+\frac{m-t}{m}) &  \text{otherwise},
   \end{cases}
\end{align} where 
\begin{itemize}
\item 
$|s_i|$ is the length of string $s_i$;
\item
$m$ is the number of matching characters from $s_1$ and $s_2$ and two characters from $s_1$ and $s_2$ are considered matching only if they are the same character and not farther than $\lfloor \frac{\max{|s_1|,|s_2|}}{2}\rfloor -1 $;
\item
 $t$ is half the number of transpositions which is defined to be the number of matching (but different sequence order) characters divided by 2.
\end{itemize}
The Jaro-Winker distance is an extension of the Jaro distance:
\begin{align}
d_{JW}(s_1,s_2)  = d_{Jaro}(s_1,s_2)+ (l\cdot p(1-d_{Jaro}(s_1,s_2))),
\end{align}
where 
\begin{itemize}
\item 
$l$ is the length of common prefix of the two string up to maximum of 4 characters;
\item
$p$ is a constant scaling factor for adjusting distance for having common prefix, the common value (applied in our case) is $p=0.1$.
\end{itemize}
Finally similarity is defined to be 
\begin{align}
S_{JW}(s_1,s_2) = 1-d_{JW}(s_1,s_2).
\end{align}
We then apply an epiweight calculation \cite{Contiero2005-ge} to the pairs in each block based on the computed similarities from the previous step, using input weight vector $w$. The epiweight  between two records $r_1,r_2$ is defined to be: 
\begin{align}
EpiWeight(r_1,r_2) = \frac{\sum_{i}p_i s(r_1^i,r_2^i)}{\sum_{i}p_i}
\end{align}
where 
\begin{itemize}
\item 
$s(r_1^i,r_2^i)$ is the similarity of two records in $i^{th}$ fields, which would be JW similarity in our case;
\item
$p_i =\frac{\log_2 (1-e_i)}{f_i}$ where $f_i$ is the average frequency of field $i$ and $e_i$ is the estimated error rate of field $i$, which is the solution of the following problem:
\begin{align}
e_i = argmin_{x_i} \sum_i(x_i-w_i)^2 \\
s.t. \, e_i \leq 1-f_i 
\end{align}
We refer interested readers to the original epiLink paper \cite{Contiero2005-ge}.

\end{itemize}
\item
 Using weights stored from the previous step, a threshold-based classification approach is then applied;  the general idea is that the more similar two records, the more likely it is that they refer to the same person. With epiweights collected for all pairs from last step, we apply a single classification threshold, $t$, to classify a pair of records $(r_i,r_j)$:
 \begin{align}
EpiWeight(r_i,r_j) \geq t \implies (r_i,r_j) \text{ is a match } \\
EpiWeight(r_i,r_j) < t \implies (r_i,r_j) \text{ is not a match } 
 \end{align}
  To determine the threshold $t$, we fit a Pareto distribution to the distribution of epiweights and calculate a quantile on the fitted model using gpdEst function in R \cite{Sariyar2010-gi}. The threshold for exceedances used to fit the distribution is determined by finding extreme quantile $q$ (from the input) of  epiweight collections. The 95\% quantile of the exceeding weights is then used as the threshold. The matches having epiweight greater than the quantile are kept and considered as a match.

\end{enumerate}

\item
\textbf{Algorithm 2: Parameter tuning for weight $w$ and quantile $q$}

\textit{Input: } Participant data and sampled 1000 contacts

\textit{Output: } Optimal weight vector $w$ and quantile $q$
\begin{enumerate}
\item
To generate the input we randomly sample 1000 contacts and link to all participants using the same blocking criteria as in algorithm 1.
\item
1000 random weights are then sampled from 7 dimensional simplex and with each weight we create 7 different exceedance quantile cutoffs $q$, ranging from 0.92 to 0.98 and spaced by 0.01, resulting in 7000 different choices of hyperparameters. 
\item
Sampled parameters are used as input in algorithm 1 to generate 7000 different sets of matched pairs.
\item
One parameter out of the candidate pool is then sampled and its matched result is manually curated for assessment of performance. To achieve a balance between too many false positives and too few true positives, we choose parameters which result in lowest false positive rate subject to at least 85\% true positive rate.
\end{enumerate}

\item

\textbf{Algorithm 3: meta-algorithm for matching pairs}

\textit{Input: } Cleaned participants and contacts data sets

\textit{Output: } Final matched pairs
  
  \begin{enumerate}

\item
  We apply Algorithm 2 to select the best parameter for the sampled dataset
  \item
   Algorithm 1 is applied to the entire dataset, with best-performing hyperparameters from the previous step.
   \item
  The matched contacts are then taken out from contact list and we create links between participants and unmatched contacts, without blocking on key fields this time.
  \item
  The same matching procedure described in steps 2-4 in Algorithm 1 is then applied with same weight vector $w$, quantile cutoff $q$. Matched pairs from this step are then combined with those obtained from step 2.
\item
   Post-processing of matched pairs:
   
Matches produced by the linking and matching algorithm are then subject to the following post-processing procedures.

We first define 4 different sets of properties:
   \begin{enumerate}
\item
Good name, defined as having average similarities of name above 0.9 and both contact and participant with at least 2 name fields.
\item
Very good name, defined as having average similarities of names above 0.95 with at least 2 name fields.
%
\item
 Good age, defined as matched contact having a valid age, and, if the participant is younger than 15, age of contact within 5 years of that of participant. The difference is relaxed to 10 years when participants are older than 15.
\item
 Good village, defined as a similarity of villages  above 0.9
    \end{enumerate}

The following three types of pairs are then removed:
   \begin{enumerate}
\item
Neither good name nor good village
\item
Neither very good name nor good age
\item
Neither good age nor good village

   \end{enumerate}

\end{enumerate}

\end{itemize}

\subsubsection{Evaluation of matched results:}

The lack of ground truth on whether a given pair is a match limits our choices of metrics for evaluating performance in matching. However, under an assumption that post-processed pairs are true matches, we used the following evaluation metrics for assessing the quality of raw census data and the performance of our linkage algorithm:
\begin{enumerate}
\item 
Percent of named contacts who do not have their age information provided
\item
Percent of named contacts who either are missing village information or live in a village outside the community

\item
Proportion of named contacts who were linked by the algorithm, which would be 100\% if the baseline census enumeration were 100\% complete, every named contact lived in a village located within the boundaries of the study community, and the matching algorithm was perfect. 
\item
 Proportion of named contacts from a village within the community who were linked by the algorithm, which would be 100\% if  the baseline census enumeration were 100\% complete, reporting of village of every named contact were 100\% accurate, and the matching algorithm was perfect.

\item
 Proportion of edges across different households, which measures the network connectedness using family as a unit instead of individual participants.  If the above conditions in 2 held and every individual formed their own household, this proportion would be 100\%, while if every participant were from the same household it would be 0\%.




\end{enumerate}

\subsubsection{Building and visualization of social networks:}

The resulting contacts are then used to build social networks, with both undirected and directed edges, using igraph package in R, with nodes being participants and edges being final matched pairs \cite{igraph}. 
 
We used Gephi with the ForceAtlas2 algorithm to visualize constructed networks \cite{Jacomy2014-vb}. In addition, the following network statistics were computed using igraph package \cite{igraph,Kolaczyk2009-fc,Kolaczyk2014-mg}: average degree, transitivity, reciprocity, average degree of separation, and proportion of coverage of the top connected component.

\subsection{Assortative mixing}

Constructed social networks can be used to understand and test social theories such as contagion or homophily between peers. One possible metric, assortative mixing, measures the tendency for like (or unlike) nodes in networks to be connected with each other. \cite{Newman2003-is} proposed a number of measures of assortative mixing appropriate to  various mixing types, e.g., discrete versus continuous node characteristics. In addition, as noted, for example, by \cite{Newman2003-is,Newman2002-rq}, while  natural networks such as protein regulation pathway often exhibit negative assortative mixing, social networks often have a positive mixing coefficient, i.e., ``like attracts like". 

 We used the assortativity coefficient defined in \cite{Newman2003-is} and computed it for the following eight baseline covariates: age, sex, village, education level, occupation, household wealth index (derived via principal components analysis from a household socioeconomic survey), alcohol use, and contraception use. Age was treated as a continuous scalar variable and the rest as discrete.
Survey questions for drinking and contraception use status are included in the appendix.

\section{Results}
\subsection{Population-level coverage of census enumeration and name generation}
 A total of 334,952 individuals were enumerated in the 32 communities during the household census; 110,118, 103,585, and 121,249 in East Uganda, Southwest Uganda, and Kenya respectively, closely matching country-based population projections\cite{Chami2014-dj}. Of enumerated residents, 168,720 were adults (aged $\geq 15$ years) eligible for administration of the name generator. Of these,  127,226 (73.24\%) had the name generator administered and named at least one contact. The proportion of census-enumerated adult residents who named at least one contact varied by region (82.57\%  in East Uganda, 75.31\% in Southwest Uganda, and 63.88\% in Kenya). Among enumerated adults who were stable community residents (defined as having lived in the community for at least 6 months of the prior year; N= 146,862), 115,979 (78.97\%) named at least one contact (85.34\%  in East Uganda, 81.67\% in Southwest Uganda, and 70.53\% in Kenya).

 Among the enumerated adults who named at least one contact, adults in East and Southwest Uganda communities reported more contacts on average (mean = 11.7 and 14.5 with SD 7.8 and 7.5 respectively) than Kenya communities (mean = 8.59, SD = 6.5). A higher proportion of named contacts in Kenyan communities had no age recorded (mean = 60.5\%, SD = 14.5\%), compared to Ugandan communities (mean= 3.42\%, SD=7.3\% in Southwest Uganda and mean = 7.15\%, SD=6.67\% in East Uganda, (See Table \ref{tab:census1}). Finally, the proportion of named contacts reported to live in a village located inside the enumerated community, and who therefore would be expected to match to a community resident enumerated in the baseline census, also varied substantially across communities, ranging, for example from 34\% to 88\% across communities in Southwestern Uganda.

 Regarding pre-processing steps, our algorithm corrected for on average 7.0\% (SD: 3.5\%) of names with nickname prefixes in Kenyan communities; nickname pre-processing was not performed for Ugandan communities. In Kenyan communities, pre-processing corrected for an average of 5.5\%  (SD 7.3\%)  of village names in Kenya, versus an average 8.7\% (SD 12.3\%) for villages in East Uganda and 6.3\%  (SD 10.4\%) in Southwest Uganda.

\subsection{Matched results and social networks}
\begin{table}[ht]
\centering
\begin{threeparttable}
 \resizebox{\textwidth}{!}{
\begin{tabular}{ccccccc}
\hline
Community & Participants (N)\tnote{a} & Contacts (N) & Contacts Missing Age \tnote{b} & Contacts resident in community\tnote{c} & Contacts Matched\tnote{d} & Contacts inside community matched\tnote{e} \\
\hline
\textbf{East Uganda} & & & & & & \\
\hline

Bugono & 11015 & 43612 & 1.83\% & 90.97\% & 56.78\% & 62.41\% \\
Kadama & 10085 & 28198 & 1.65\% & 86.09\% & 58.40\% & 67.83\% \\
Kameke & 11197 & 25586 & 21.18\% & 79.83\% & 57.04\% & 71.45\% \\
Kamuge & 12283 & 62347 & 1.05\% & 88.33\% & 62.07\% & 70.28\% \\
Kiyeyi & 11476 & 33989 & 12.49\% & 70.91\% & 60.32\% & 85.06\% \\
Kiyunga & 10991 & 44187 & 2.32\% & 47.28\% & 50.10\% & 100.00\% \\
Merikit & 11560 & 37415 & 9.24\% & 52.52\% & 57.20\% & 100.00\% \\
Muyembe & 12285 & 37582 & 2.50\% & 86.50\% & 55.40\% & 64.04\% \\
Nankoma & 11442 & 34125 & 6.54\% & 76.70\% & 46.64\% & 60.81\% \\
Nsiinze & 9834 & 16219 & 12.74\% & 66.05\% & 43.57\% & 65.97\% \\

\hline
\textbf{Southwest Uganda} & & & & & & \\
\hline

Bugamba & 12457 & 92563 & 2.58\% & 80.95\% & 48.20\% & 59.54\% \\
Kazo & 13018 & 32931 & 0.76\% & 72.20\% & 55.49\% & 76.85\% \\
Kitwe & 10165 & 44924 & 0.33\% & 37.24\% & 49.73\% & 100.00\% \\
Mitooma & 10287 & 52945 & 0.45\% & 49.86\% & 54.73\% & 100.00\% \\
Nsiika & 10575 & 52291 & 3.87\% & 87.77\% & 42.33\% & 48.23\% \\
Nyamuyanja & 9398 & 29396 & 24.13\% & 68.53\% & 45.44\% & 66.31\% \\
Rubaare & 9529 & 37510 & 0.70\% & 46.50\% & 48.23\% & 100.00\% \\
Rugazi & 9778 & 57103 & 0.82\% & 34.12\% & 43.43\% & 100.00\% \\
Ruhoko & 9512 & 30424 & 0.35\% & 55.24\% & 56.20\% & 100.00\% \\
Rwashamire & 9325 & 29734 & 0.26\% & 44.93\% & 45.21\% & 100.00\% \\
\hline
\textbf{Kenya} & & & & & & \\
\hline

Bware & 8673 & 14974 & 69.57\% & 87.00\% & 22.66\% & 26.05\% \\
kisegi & 11370 & 26041 & 60.82\% & 90.59\% & 25.41\% & 28.05\% \\
Kitare & 9630 & 27505 & 60.20\% & 80.76\% & 27.29\% & 33.79\% \\
Magunga & 11075 & 25201 & 49.22\% & 88.14\% & 30.71\% & 34.84\% \\
Nyamrisra & 8630 & 10119 & 40.97\% & 66.18\% & 42.01\% & 63.48\% \\
Nyatoto & 11497 & 11328 & 36.49\% & 84.87\% & 46.70\% & 55.02\% \\
Ogongo & 11781 & 24338 & 44.03\% & 75.08\% & 41.57\% & 55.37\% \\
Ongo & 10431 & 21998 & 73.50\% & 73.72\% & 17.72\% & 24.04\% \\
Othoro & 9626 & 18360 & 73.33\% & 82.78\% & 19.61\% & 23.69\% \\
Sena & 8642 & 14042 & 72.39\% & 79.35\% & 21.84\% & 27.53\% \\
Sibuoche & 11488 & 18039 & 79.16\% & 81.28\% & 17.81\% & 21.91\% \\
Tom Mboya & 8406 & 18237 & 66.62\% & 86.08\% & 26.48\% & 30.77\% \\
\hline
\end{tabular}

}

\begin{tablenotes}\footnotesize
\item [a] Number of residents (both adult and children) enumerated during census.
\item [b] Percent of named contacts who did not have their age information provided.
\item [c] Percent of named contacts who were reported to live in a village inside the community
\item [d] Percent of contacts successfully matched to an enumerated study participant
\item [e] Percent of named contacts reported to live in a village inside the community successfully matched 

to an enumerated study participant  by the algorithm.

\end{tablenotes}
  \caption{
  Summary of census enumeration data in the SEARCH study with data quality and matching metrics}
    \label{tab:census1}
\end{threeparttable}

\end{table}

We matched named contacts within each community using our proposed algorithm, with results of  contact matching listed in Table \ref{tab:census1}. 

The percentage of named contacts who were successfully matched varied greatly across geographical regions, both before and after accounting for variability in the proportion of named contacts who lived within the enumerated community. On average, 28.42\% of all named contacts   (SD $= 9.99\%$) and 35.38\% of named contacts reported to live within the community (SD = 14.30\%) were successfully matched in Kenya; $48.70\%$ (SD $= 5.16\%$) of all contacts and 85.09\% (SD = 20.45\%) of within-community contacts were successfully matched in Southwest Uganda; and $54.7\%$ (SD $= 5.16\%$) of all contacts and 74.79\% (SD = 14.9\%) of within-community contacts were successfully matched in East Uganda. Proportion of named contacts who were successfully matched was closely correlated with proportion of named contacts with non-missing age (correlation 0.90).

Summary statistics of the resulting social networks, restricted to adult (aged $\geq 15$ years) residents are shown in Table \ref{tab:full_network}. Kenyan communities had less dense networks, with an average degree of 1.6, in contrast to an average degree of 5.69 and 5.84 in East and Southwest Ugandan communities respectively. Overall our network characteristics corroborated patterns in \cite{Tsai_Perkins2018-is} where an average degree of 9.1 and 8·6 for female and male networks in Uganda were reported.

Transitivity and reciprocity were more similar across regions; Kenyan communities had average transitivity 0.12 (SD = 0.04) and reciprocity 0.16 (SD = 0.03) compared to average transitivity 0.14 (SD = 0.03) and reciprocity 0.22 (SD = 0.03) in Uganda. Finally, networks for Kenyan communities were sparser than those for Uganda, as expected given the lower numbers of matched contacts in Kenya.  In addition, largest connected components in Uganda, normalized by sizes of graphs, were larger than those in Kenya.

 Corresponding summary statistics of subnetworks among ``stable" adult residents, defined as adult residents who reported living in the community for more than 6 months of the past year, are provided in
Table \ref{tab:stable_network}. Similar trends held for subgraphs among stable residents, with Kenyan communities less dense and connected with respect to every metric, compared to Ugandan communities.

\begin{table}
\centering
\begin{threeparttable}

 \resizebox{\textwidth}{!}{
 \begin{tabular}{ccccccccc}
  \hline
Community & Nodes\tnote{a} & Edges\tnote{b} (Across household percent) & Average Degrees & Transitivity & Reciprocity & Avg. Path Length\tnote{c} & Top CC Coverage\tnote{d} \\
\hline
\textbf{East Uganda} & & & & & & & & \\
\hline
Bugono & 5035 & 18129 (81.86\%) & 7.20 & 0.10 & 0.26 & 4.66 & 87.75\% \\
Kadama & 4515 & 11881 (81.21\%) & 5.26 & 0.16 & 0.26 & 5.84 & 85.69\% \\
Kameke & 5177 & 11520 (81.98\%) & 4.45 & 0.09 & 0.19 & 5.52 & 83.60\% \\
Kamuge & 5434 & 28567 (87.15\%) & 10.51 & 0.16 & 0.25 & 4.59 & 94.39\% \\
Kiyeyi & 5308 & 15543 (87.82\%) & 5.86 & 0.13 & 0.19 & 5.15& 88.58\% \\
kiyunga & 5081 & 16234 (83.81\%) & 6.39 & 0.10 & 0.21 & 4.90 & 92.64\% \\
Merikit & 5648 & 16142 (88.42\%) & 5.72 & 0.15 & 0.20 & 5.23 & 88.83\% \\
Muyembe & 6740 & 14428 (79.23\%) & 4.28 & 0.22 & 0.22 & 6.13 & 76.59\% \\
Nankoma & 5069 & 11868 (84.31\%) & 4.68 & 0.12 & 0.22 & 5.77 & 85.78\% \\
Nsiinze & 4629 & 5916 (78.52\%) & 2.56 & 0.10 & 0.21 & 7.84 & 74.88\% \\
\hline
\textbf{Southwest Uganda} & & & & & & & & \\
\hline
Bugamba & 6630 & 31779 (87.89\%) & 9.59 & 0.15 & 0.22 & 4.72 & 92.84\% \\
Kazo & 6829 & 13798 (75.79\%) & 4.04 & 0.14 & 0.23 & 6.79 & 82.00\% \\
kitwe & 4942 & 16294  (82.46\%) & 6.59 & 0.14 & 0.25 & 5.28 & 91.44\% \\
Mitooma & 5615 & 20662 (82.67\%) & 7.36 & 0.13 & 0.25 & 4.85 & 94.18\% \\
Nsiika & 5450 & 16135 (85.57\%) & 5.92 & 0.15 & 0.21 & 6.39 & 86.17\% \\
Nyamuyanja & 4801 & 10161 (80.74\%) & 4.23 & 0.12 & 0.22 & 6.06 & 84.02\% \\
Rubaare & 5055 & 13241 (78.32\%) & 5.24 & 0.15 & 0.26 & 5.80 & 87.40\% \\
Rugazi & 5024 & 17086 (83.82\%) & 6.80 & 0.16 & 0.26 & 5.06 & 91.94\% \\
Ruhoko & 5412 & 12883 (77.91\%) & 4.76 & 0.14 & 0.21 & 5.98& 84.16\% \\
Rwashamire & 5127 & 9926 (75.05\%) & 3.87 & 0.15 & 0.23 & 6.45 & 83.79\% \\
\hline
\textbf{Kenya} & & & & & & & & \\
\hline
Bware & 4634 & 2349 (58.71\%) & 1.01 & 0.18 & 0.20 & 15.54 & 26.39\% \\
kisegi & 5738 & 5202 (75.95\%) & 1.81 & 0.16 & 0.20 & 7.93 & 53.29\% \\
Kitare & 4849 & 5784 (80.15\%) & 2.39 & 0.11 & 0.17 & 7.11 & 67.42\% \\
Magunga & 5716 & 6212 (77.85\%) & 2.17 & 0.11 & 0.18 & 7.56 & 63.61\% \\
Nyamrisra & 4449 & 3385 (69.54\%) & 1.52 & 0.07 & 0.21 & 11.23 & 55.20\% \\
Nyatoto & 5936 & 4259 (72.20\%) & 1.43 & 0.07 & 0.15 & 11.24 & 50.40\% \\
Ogongo & 6276 & 8104 (77.75\%) & 2.58 & 0.10 & 0.20 & 7.28 & 72.23\% \\
Ongo & 5296 & 3742 (75.17\%) & 1.41 & 0.14 & 0.15 & 8.5 & 44.94\% \\
Othoro & 4758 & 2770 (73.61\%) & 1.16 & 0.16 & 0.20 & 9.13 & 34.91\% \\
Sena & 4639 & 2469 (69.87\%) & 1.06 & 0.08 & 0.11 & 10.01 & 34.12\% \\
Sibuoche & 5603 & 2605 (65.49\%) & 0.93 & 0.15 & 0.19 & 11.74 & 27.63\% \\
Tom Mboya & 4613 & 3891 (76.30\%) & 1.69 & 0.10 & 0.14 & 8.10 & 53.89\% \\
\hline

\end{tabular}

}

\begin{tablenotes}\footnotesize
\item [a] Number of adult residents enumerated during census 
\item [b] Edges in undirected graph
\item [c] For paths between connected vertices
\item [d] Proportion of nodes in the largest connected component of the network
\end{tablenotes}

\end{threeparttable}

  \caption{Summary statistics of adult social networks for 32 parishes in the SEARCH study}
    \label{tab:full_network}

\end{table}

\begin{figure}[ht!]
\centering
\includegraphics[scale=0.3]{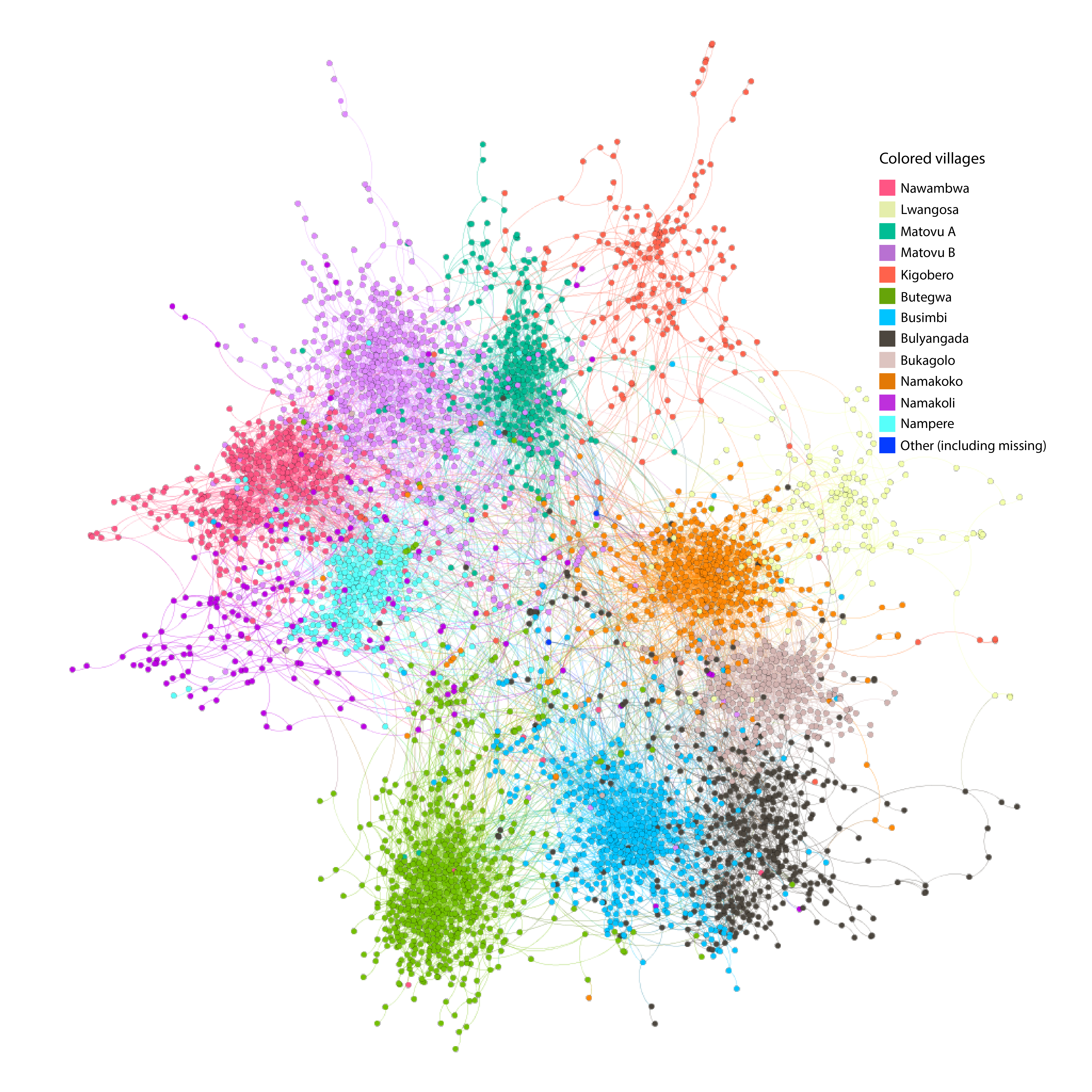}
\caption{Largest connected component of constructed social network in Nankoma  where each color represents a unique village}
\label{nankoma_network}
\end{figure}

As illustration, Figure \ref{nankoma_network} shows the largest connected component in Nankoma community with each color representing a unique village, using a ForceAtlas2 layout\cite{Jacomy2014-vb}. 
 
\subsection{Mixing patterns}

\begin{figure}[ht!]
\centering
\includegraphics[scale=0.1]{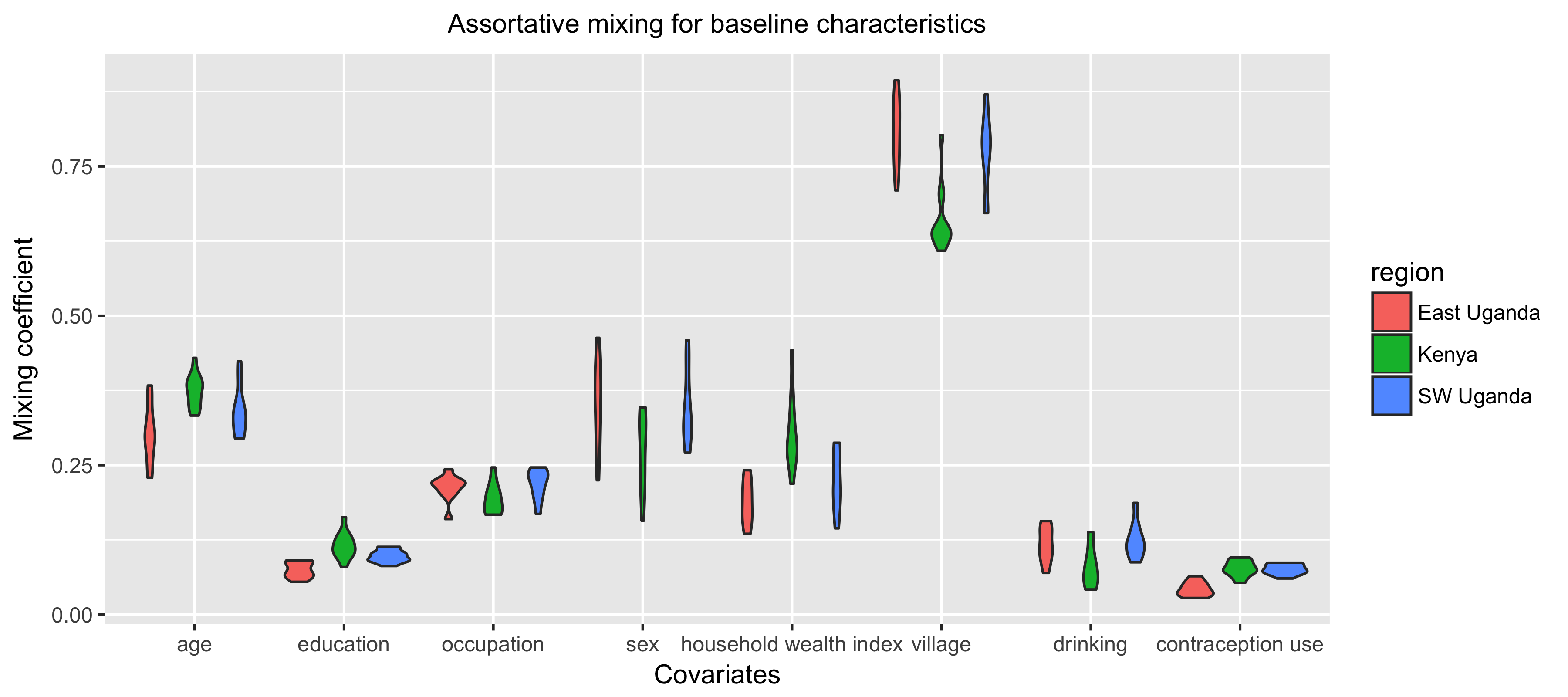}
\caption{Assortative mixing pattern of covariates across different communities}
\label{assortative_mix_patterns}
\end{figure}

Assortative mixing coefficients were calculated for all communities in eight different domains and plotted in a violin plot (Figure \ref{assortative_mix_patterns}). 

Overall we observed positive associations among community members with similar covariates, with reasonable consistency across regions. Among the eight covariates considered, village, sex, age, and household wealth index had high assortative mixing, with averages from 0.4-0.7. In comparison, those reported in \cite{Newman2003-is} and \cite{Chow2016-mq} for the networks deemed as assortative were generally in the range of 0.2-0.6.

\section{Discussion}

 We constructed 32 large-scale sociocentric social networks in rural Sub-Saharan Africa using a record-linkage algorithm applied to link census enumerated individuals with contacts generated using  a five question name generator. To the best of our knowledge, this is the largest set of sociocentric networks in Sub-Saharan Africa described in the literature, incorporating 170,028 adult residents and 362,965 relationships in 32 communities in rural Uganda and Kenya. Success at matching named contacts living within the community ranged from 22\% to 100\% across communities. As one of the first detailed descriptions of the network building process in the absence of real time linkage at time of name generation in rural resource-limited settings, this paper illustrates one process for generating large scale sociocentric  networks in prototypic Sub-Saharan rural communities that may enable the pursuit of novel questions and insights into public health. As a proof of concept exercise, we also demonstrated the mixing or homophily pattern for a number of covariates in the constructed networks.

The combination of scale (total number of residents in the networks we mapped) and limited resources available during baseline census enumeration and name generation resulted in a dataset that that was not amenable to commonly used approaches for network construction. Specifically, current literature on existing record linkage algorithms is largely focused on their application to manually curated census datasets in the U.S. or other  resources-rich regions  \cite{Sariyar2010-gi,Contiero2005-ge,Christen2012-bm,Randall2013-mb}.  We instead developed a novel approach to network construction that builds on and extends existing matching algorithms and incorporates assessment of matching quality. To optimize the trade-off between scalability and number of true matches, we modified the traditional blocking criteria to incorporate two-step blocking. We incorporated a widely-applied blocking step in our algorithm \cite{Christen2012-bm,Contiero2005-ge} and then introduced a separate unblocking matching step for unmatched records, as documented in record linkage reviews such as\cite{Christen2012-bm}. In addition, to minimize the impact of varying raw data quality, we proposed a comprehensive set of pre-processing and post-processing rules. While the algorithm still largely operates within the classical record linkage framework, it offers practitioners a complete procedure to clean, match, and analyze survey-based network data. It thus contributes to the  literature applying explicit record linkage algorithms to network data, and may facilitate network construction for future applied network analyses.

Both our approach to network construction and the resulting networks we report are subject to several limitations. Complete census enumeration of entire communities and administration of a name generator questionnaire to most residents is resource-intensive and the challenges of data collection in the field impacts the completeness and accuracy of the networks. The household-based census enumerated a similar number of residents as projected by recent official government census \cite{Chamie2016-sd}; however some individuals still may have been missed. Name matching is challenged by the following observations in this cultural context: cultural variations of names may not have been captured in our algorithm; individuals may go by multiple different names; and, persons may have used alternate names to protect confidentiality. We also did not explicitly ask about the sex of contacts and thus relied on imputation for this information. Age estimation can also be challenging in a region where birth certificates are not issued. In order to mitigate the challenges associated with data collection in the field, the name generator was administered by local field staff who spoke the same language as those interviewed, and community sensitization efforts in collaboration with local leaders were undertaken prior to the census in order to prepare the community for personal questions. Real time quality control using a community census for validation at the time of administration of the name generator would likely reduce these potential errors and could improve matching in future work.  
Finally, the weight parameter selection during network construction still requires manual inspection, and our matched results have not been externally verified.

The matched ratio of a constructed network provides a context for interpreting downstream analyses. However, further toolkits are needed to quantify the effect of both missing nodes (due to incompletely enumerated residents) and missing edges (due to missing or erroneous contact information leading to missed or incorrect matches) on downstream analyses. For instance, differences in network summary statistics and estimates of network assortativity seen across regions could be attributed to a greater proportion of true edges that remained undetected by our algorithm in Kenya compared to Uganda.   

The constructed networks in the SEARCH study offer a platform for improving understanding of the role of networks in health outcomes, and new opportunities for novel and tailored interventions. The algorithm for social network construction from census enumeration data described here is scalable and includes performance metrics that are easily computed for network quality assessment. Including network construction in population-based research studies opens the door to a wide range of  analyses and may enrich our understanding of the role social networks play in communicable and non-communicable diseases. Network data can also be combined with phylogenetic data and geospatial data to provide new, multi-dimensional insights about disease transmission and diffusion of health-related behaviors.

\section{Appendix}

\subsection{Name generators}
The following five different name generators were used to ask information about participants' social network data:

The following information was provided with before asking specific contact questions:
For the next 5 questions, I am going to ask you about specific relationships you may have. You may name up to six people in response to each question. It is acceptable if you repeat names for different questions.  Please only name people aged 15 years or older.  I will ask you to tell me their names. Please do not name anyone who has passed away. Tell me their full names, including any nicknames.

\subsubsection{Money contacts}
\begin{itemize}
\item 
Over the last 12 months, with whom have you usually discussed any kind of money matters? Examples of money matters might include school fees; employment; giving, receiving, or paying loans; starting businesses; financing for big events; or other financial issues. 
\item
Please remember that you may include the names of people that you have named in response to previous questions.
\item
What is your contact's age?
\end{itemize}

\subsubsection{Health contacts}
\begin{itemize}
\item 
Over the past 12 months, with whom have you usually discussed any kind of health issue? Examples of health issues might include topics like your child’s health, family planning, nutrition, HIV, mental health, immunizations, sanitation methods, alcohol abuse or other issues.
\item
Please remember that you may include the names of people that you have named in response to previous questions.
\item
What is your contact's age?
\end{itemize}

\subsubsection{Emotional support contacts}
\begin{itemize}
\item 
Over the past 12 months, to whom have you gone to receive emotional support? Examples of emotional support might include talking about either positive or negative topics such as deaths, marriages, births, loss of a job, or other topics of emotional importance for you. 
\item
Please remember that you may include the names of people that you have named in response to previous questions.
\item
What is your contact's age?
\end{itemize}

\subsubsection{Free time contacts}
\begin{itemize}
\item 
Over the past 12 months, with whom have you usually spent time for your leisure, enjoyment, relaxation, at parties, attending trainings together of your choice, watching sports games, taking alcohol together, weaving mats, or whenever you have made time for yourself (free time)? 
\item
Please remember that you may include the names of people that you have named in response to previous questions.
\item
What is your contact's age?
\end{itemize}

\subsubsection{Food contacts}
\begin{itemize}
\item 
 Over the past 12 months, with whom have you shared, borrowed, received, or exchanged any food? 

For this question only, please do not name people who regularly stay at your household.
\item
Please remember that you may include the names of people that you have named in response to previous questions.
\item
What is your contact's age?

\end{itemize}
\subsection{Summary statistics of stable residents' social networks}
\begin{table}[ht!]
\centering
\begin{threeparttable}
 \resizebox{\textwidth}{!}{\begin{tabular}{cccccccc}
\hline
Community   &  Nodes &  Edges\tnote{a} & Average Deg & Transitivity & Reciprocity & Avg Path Length\tnote{b} & Top CC Coverage\tnote{c} \\
\hline
\textbf{East Uganda} &&&&&&&\\
\hline
Bugono & 3897 & 15001 & 7.70 & 0.10 & 0.23 & 4.59 & 93.76\% \\
Kadama & 4269 & 11287 & 5.29 & 0.16 & 0.25 & 5.81 & 86.39\% \\
Kameke & 4915 & 10831 & 4.41 & 0.10 & 0.20 & 5.64 & 84.76\% \\
Kamuge & 5000 & 26401 & 10.56 & 0.17 & 0.25 & 4.57 & 94.90\% \\
Kitare & 4250 & 4983 & 2.34 & 0.12 & 0.17 & 7.21 & 68.05\% \\
Kiyeyi & 4940 & 14513 & 5.88 & 0.14 & 0.19 & 5.16 & 89.39\% \\
Kiyunga & 4769 & 15133 & 6.35 & 0.10 & 0.21 & 4.91 & 93.08\% \\
Merikit & 5150 & 14621 & 5.68 & 0.15 & 0.19 & 5.27 & 90.08\% \\
Muyembe & 5242 & 12552 & 4.79 & 0.23 & 0.22 & 6.11 & 84.95\% \\
Nankoma & 4784 & 11109 & 4.64 & 0.12 & 0.22 & 5.78 & 86.50\% \\
Nsiinze & 4360 & 5532 & 2.54 & 0.10 & 0.20 & 7.91 & 75.41\% \\
\hline
\textbf{Southwest Uganda} &&&&&&&\\
\hline
Bugamba & 5640 & 27931 & 9.90 & 0.15 & 0.25 & 4.70 & 95.87\% \\
Kazo & 5609 & 11766 & 4.20 & 0.14 & 0.23 & 6.86 & 85.43\% \\
kitwe & 4377 & 14773 & 6.75 & 0.14 & 0.26 & 5.25 & 93.85\% \\
Mitooma & 4982 & 18357 & 7.37 & 0.14 & 0.25 & 4.87 & 95.24\% \\
Nsiika & 4987 & 14723 & 5.90 & 0.16 & 0.24 & 6.30 & 87.59\% \\
Nyamuyanja & 3952 & 8949 & 4.53 & 0.13 & 0.23 & 5.96 & 88.82\% \\
Rubaare & 4534 & 12267 & 5.41 & 0.15 & 0.25 & 5.80 & 90.27\% \\
Rugazi & 4397 & 15221 & 6.92 & 0.16 & 0.26 & 5.03 & 93.22\% \\
Ruhoko & 4180 & 10522 & 5.03 & 0.14 & 0.23 & 5.98 & 89.14\% \\
Rwashamire & 4346 & 8747 & 4.03 & 0.15 & 0.23 & 6.40 & 86.75\% \\
\hline
\textbf{Kenya} &&&&&&&\\
\hline
Bware & 3727 & 1904 & 1.02 & 0.17 & 0.20 & 16.30 & 27.90\% \\
Kisegi & 4654 & 4133 & 1.78 & 0.16 & 0.18 & 8.09 & 52.81\% \\
Magunga & 4672 & 5115 & 2.19 & 0.12 & 0.16 & 7.77 & 65.92\% \\
Nyamrisra & 3962 & 2909 & 1.47 & 0.07 & 0.22 & 11.91 & 53.53\% \\
Nyatoto & 5172 & 3770 & 1.40 & 0.06 & 0.19 & 12.79 & 50.27\% \\
Ogongo & 5318 & 6771 & 2.55 & 0.10 & 0.19 & 7.66 & 73.52\% \\
Ongo & 4368	& 3070 & 1.41 & 0.14 &  0.15 & 8.80 & 45.79\% \\
Othoro & 4042 & 2457 & 1.22 & 0.16 & 0.20 & 9.24 & 37.56\% \\
Sena & 3718 & 1950 & 1.05 & 0.08 & 0.14 & 10.56 & 33.43\% \\
Sibuoche & 4718	 & 2199 & 0.93	 & 0.16	& 0.20	 & 12.30	 & 27.87\% \\
Tom Mboya & 3744 & 3095 & 1.65 & 0.11 & 0.14 & 8.57 & 54.86\% \\
\hline

\end{tabular}
}

\begin{tablenotes}\footnotesize
\item [a] Edges in undirected graph
\item [b] We only consider paths between connected nodes
\item [c] Proportion of nodes in the largest connected component of the network
\end{tablenotes}

\end{threeparttable}

  \caption{Summary statistics of stable adult social networks for 32 parishes in the SEARCH study}
    \label{tab:stable_network}

\end{table}

\subsection{Measures of risky behavior:}
 
For alcohol use, participants were asked to report their consumption of alcohol: “Do you drink alcohol?” Responses were encoded in four different categories: No, Yes, Refused to answer,  and Skipped. If participants answered “Yes”, they would be further asked to answer “How many days in a month do you drink alcohol?” and “How many drinks containing alcohol do you have on a typical day, when drinking?”
 
For contraceptive use, participants were asked: “Are you or your partner currently using contraception?” and again results were encoded in No, Yes, Refused to answer, and Skipped categories. If participants answered “Yes”, they would answer the question “What contraceptive are you or your partner currently using?”
 
For the scope of our analysis, we encode Yes and No to be the binary outcomes, and both Refused to answer, and Skipped were grouped under “Missing” category, which was treated as node censoring in subsequent analysis.

\nolinenumbers

\bibliography{SN}

\begin{thebibliography}{}

\bibitem[Amirkhanian, 2014]{Amirkhanian2014-xy}
Amirkhanian, Y.~A. (2014).
\newblock Social networks, sexual networks and {HIV} risk in men who have sex
  with men.
\newblock {\em Curr. HIV/AIDS Rep.}, 11(1):81--92.

\bibitem[Bollen et~al., 2011]{Bollen2011-eo}
Bollen, J., Goncalves, B., Ruan, G., and Mao, H. (2011).
\newblock Happiness is assortative in online social networks.

\bibitem[Chami et~al., 2014]{Chami2014-dj}
Chami, G.~F., Ahnert, S.~E., Voors, M.~J., and Kontoleon, A.~A. (2014).
\newblock Social network analysis predicts health behaviours and self-reported
  health in african villages.
\newblock {\em PLoS One}, 9(7):e103500.

\bibitem[Chamie et~al., 2016]{Chamie2016-sd}
Chamie, G., Clark, T.~D., Kabami, J., Kadede, K., Ssemmondo, E., Steinfeld, R.,
  Lavoy, G., Kwarisiima, D., Sang, N., Jain, V., Thirumurthy, H., Liegler, T.,
  Balzer, L.~B., Petersen, M.~L., Cohen, C.~R., Bukusi, E.~A., Kamya, M.~R.,
  Havlir, D.~V., and Charlebois, E.~D. (2016).
\newblock A hybrid mobile approach for population-wide {HIV} testing in rural
  east africa: an observational study.
\newblock {\em Lancet HIV}, 3(3):e111--9.

\bibitem[Choice and Adolescents, 1998]{Choice1998-ny}
Choice, F. and Adolescents, F. I.~o. (1998).
\newblock {LEAD} {ARTICLE}.
\newblock {\em Prev. Med.}, 27:645--656.

\bibitem[Chow et~al., 2016]{Chow2016-mq}
Chow, E. P.~F., Read, T. R.~H., Law, M.~G., Chen, M.~Y., Bradshaw, C.~S., and
  Fairley, C.~K. (2016).
\newblock Assortative sexual mixing patterns in male?female and male?male
  partnerships in melbourne, australia: implications for {HIV} and sexually
  transmissible infection transmission.
\newblock {\em Sex. Health}.

\bibitem[Christakis and Fowler, 2007]{Christakis2007-ir}
Christakis, N.~A. and Fowler, J.~H. (2007).
\newblock The spread of obesity in a large social network over 32 years.
\newblock {\em N. Engl. J. Med.}, 357(4):370--379.

\bibitem[Christakis and Fowler, 2008]{Christakis2008-rc}
Christakis, N.~A. and Fowler, J.~H. (2008).
\newblock The collective dynamics of smoking in a large social network.
\newblock {\em N. Engl. J. Med.}, 358(21):2249--2258.

\bibitem[Christakis and Fowler, 2013]{Christakis2013-tk}
Christakis, N.~A. and Fowler, J.~H. (2013).
\newblock Social contagion theory: examining dynamic social networks and human
  behavior.
\newblock {\em Stat. Med.}, 32(4):556--577.

\bibitem[Christen, 2012]{Christen2012-bm}
Christen, P. (2012).
\newblock {\em Data Matching: Concepts and Techniques for Record Linkage,
  Entity Resolution, and Duplicate Detection}.
\newblock Data-Centric Systems and Applications. Springer Berlin Heidelberg.

\bibitem[Christley et~al., 2005]{Christley2005-yy}
Christley, R.~M., Pinchbeck, G.~L., Bowers, R.~G., Clancy, D., French, N.~P.,
  Bennett, R., and Turner, J. (2005).
\newblock Infection in social networks: using network analysis to identify
  high-risk individuals.
\newblock {\em Am. J. Epidemiol.}, 162(10):1024--1031.

\bibitem[Cohen-Cole and Fletcher, 2008]{Cohen-Cole2008-gw}
Cohen-Cole, E. and Fletcher, J.~M. (2008).
\newblock Is obesity contagious? social networks vs. environmental factors in
  the obesity epidemic.
\newblock {\em J. Health Econ.}, 27(5):1382--1387.

\bibitem[Contiero et~al., 2005]{Contiero2005-ge}
Contiero, P., Tittarelli, A., Tagliabue, G., Maghini, A., Fabiano, S.,
  Crosignani, P., and Tessandori, R. (2005).
\newblock The {EpiLink} record linkage software: presentation and results of
  linkage test on cancer registry files.
\newblock {\em Methods Inf. Med.}, 44(1):66--71.

\bibitem[Danon et~al., 2011]{Danon2011-jl}
Danon, L., Ford, A.~P., House, T., Jewell, C.~P., Keeling, M.~J., Roberts,
  G.~O., Ross, J.~V., and Vernon, M.~C. (2011).
\newblock Networks and the epidemiology of infectious disease.
\newblock {\em Interdiscip. Perspect. Infect. Dis.}, 2011:284909.

\bibitem[Eubank et~al., 2004]{Eubank2004-bb}
Eubank, S., Guclu, H., Kumar, V. S.~A., Marathe, M.~V., Srinivasan, A.,
  Toroczkai, Z., and Wang, N. (2004).
\newblock Modelling disease outbreaks in realistic urban social networks.
\newblock {\em Nature}, 429(6988):180--184.

\bibitem[Fowler and Christakis, 2008a]{Fowler2008-qb}
Fowler, J.~H. and Christakis, N.~A. (2008a).
\newblock Dynamic spread of happiness in a large social network: longitudinal
  analysis over 20 years in the framingham heart study.
\newblock {\em BMJ}, 337:a2338.

\bibitem[Fowler and Christakis, 2008b]{Fowler2008-rm}
Fowler, J.~H. and Christakis, N.~A. (2008b).
\newblock Estimating peer effects on health in social networks: a response to
  {Cohen-Cole} and fletcher; and trogdon, nonnemaker, and pais.
\newblock {\em J. Health Econ.}, 27(5):1400--1405.

\bibitem[Gile and Handcock, 2015]{Gile2015-gm}
Gile, K.~J. and Handcock, M.~S. (2015).
\newblock Network {Model-Assisted} inference from {Respondent-Driven} sampling
  data.
\newblock {\em J. R. Stat. Soc. Ser. A Stat. Soc.}, 178(3):619--639.

\bibitem[{Gábor Csárdi et al.}, 2017]{igraph}
{Gábor Csárdi et al.} (2017).
\newblock {\em {R} Package igraph}.

\bibitem[Helleringer et~al., 2009]{Likoma}
Helleringer, S., Kohler, H.-P., Chimbiri, A., Chatonda, P., and Mkandawire, J.
  (2009).
\newblock The likoma network study: Context, data collection, and initial
  results.
\newblock {\em Demogr. Res.}, 21:427--468.

\bibitem[Jacomy et~al., 2014]{Jacomy2014-vb}
Jacomy, M., Venturini, T., Heymann, S., and Bastian, M. (2014).
\newblock {ForceAtlas2}, a continuous graph layout algorithm for handy network
  visualization designed for the gephi software.
\newblock {\em PLoS One}, 9(6):e98679.

\bibitem[Keeling and Eames, 2005]{Keeling2005-tz}
Keeling, M.~J. and Eames, K. T.~D. (2005).
\newblock Networks and epidemic models.
\newblock {\em J. R. Soc. Interface}, 2(4):295--307.

\bibitem[Kelly et~al., 2014]{Kelly2014-vw}
Kelly, L., Patel, S.~A., Narayan, K. M.~V., Prabhakaran, D., and Cunningham,
  S.~A. (2014).
\newblock Measuring social networks for medical research in lower-income
  settings.
\newblock {\em PLoS One}, 9(8):e105161.

\bibitem[Klovdahl, 1985]{Klovdahl1985-ta}
Klovdahl, A.~S. (1985).
\newblock Social networks and the spread of infectious diseases: the {AIDS}
  example.
\newblock {\em Soc. Sci. Med.}, 21(11):1203--1216.

\bibitem[Klovdahl et~al., 1994]{Klovdahl1994-zz}
Klovdahl, A.~S., Potterat, J.~J., Woodhouse, D.~E., Muth, J.~B., Muth, S.~Q.,
  and Darrow, W.~W. (1994).
\newblock Social networks and infectious disease: the colorado springs study.
\newblock {\em Soc. Sci. Med.}, 38(1):79--88.

\bibitem[Kolaczyk, 2009]{Kolaczyk2009-fc}
Kolaczyk, E.~D. (2009).
\newblock {\em Statistical Analysis of Network Data: Methods and Models}.
\newblock Springer Series in Statistics. Springer New York.

\bibitem[Kolaczyk and Cs{\'a}rdi, 2014]{Kolaczyk2014-mg}
Kolaczyk, E.~D. and Cs{\'a}rdi, G. (2014).
\newblock {\em Statistical Analysis of Network Data with R:}.
\newblock Use R! Springer New York.

\bibitem[Kossinets and Watts, 2006]{Kossinets2006-rv}
Kossinets, G. and Watts, D.~J. (2006).
\newblock Empirical analysis of an evolving social network.
\newblock {\em Science}, 311(5757):88--90.

\bibitem[Kramer et~al., 2014]{Kramer2014-my}
Kramer, A. D.~I., Guillory, J.~E., and Hancock, J.~T. (2014).
\newblock Experimental evidence of massive-scale emotional contagion through
  social networks.
\newblock {\em Proc. Natl. Acad. Sci. U. S. A.}, 111(24):8788--8790.

\bibitem[Kretzschmar and Morris, 1996]{Kretzschmar1996-vc}
Kretzschmar, M. and Morris, M. (1996).
\newblock Measures of concurrency in networks and the spread of infectious
  disease.
\newblock {\em Math. Biosci.}, 133(2):165--195.

\bibitem[Latkin and Knowlton, 2015]{Latkin2015-te}
Latkin, C.~A. and Knowlton, A.~R. (2015).
\newblock Social network assessments and interventions for health behavior
  change: A critical review.
\newblock {\em Behav. Med.}, 41(3):90--97.

\bibitem[Lemieux-Mellouki et~al., 2016]{Lemieux-Mellouki2016-lz}
Lemieux-Mellouki, P., Drolet, M., Brisson, J., Franco, E.~L., Boily, M.-C.,
  Baussano, I., and Brisson, M. (2016).
\newblock Assortative mixing as a source of bias in epidemiological studies of
  sexually transmitted infections: the case of smoking and human
  papillomavirus.
\newblock {\em Epidemiol. Infect.}, 144(7):1490--1499.

\bibitem[Lin et~al., 2012]{Lin2012-jn}
Lin, H., He, N., Ding, Y., Qiu, D., Zhu, W., Liu, X., Zhang, T., and Detels, R.
  (2012).
\newblock Tracing sexual contacts of {HIV-infected} individuals in a rural
  prefecture, eastern china.
\newblock {\em BMC Public Health}, 12:533.

\bibitem[Lipsitch et~al., 2003]{Lipsitch2003-wo}
Lipsitch, M., Cohen, T., Cooper, B., Robins, J.~M., Ma, .~S., James, L.,
  Gopalakrishna, G., Chew, S.~K., Tan, . C.~C., Samore, M.~H., Fisman, D., and
  Murray, .~M. (2003).
\newblock Transmission dynamicsand controlof severe acute respiratory syndrome.
\newblock {\em Science}, 300.

\bibitem[Loucks et~al., 2006]{Loucks2006-dj}
Loucks, E.~B., Sullivan, L.~M., D'Agostino, Sr, R.~B., Larson, M.~G., Berkman,
  L.~F., and Benjamin, E.~J. (2006).
\newblock Social networks and inflammatory markers in the framingham heart
  study.
\newblock {\em J. Biosoc. Sci.}, 38(6):835--842.

\bibitem[Morris, 2004]{Morris2004-xi}
Morris, M. (2004).
\newblock The collection and analysis of social network data in nang rong,
  thailand.
\newblock In {\em Network Epidemiology}. Oxford University Press, Oxford.

\bibitem[Mulawa et~al., 2018]{Mulawa2018-vd}
Mulawa, M.~I., Yamanis, T.~J., Kajula, L.~J., Balvanz, P., and Maman, S.
  (2018).
\newblock Structural network position and performance of health leaders within
  an {HIV} prevention trial.
\newblock {\em AIDS Behav.}, 22(9):3033--3043.

\bibitem[Newman, 2002]{Newman2002-rq}
Newman, M. E.~J. (2002).
\newblock Spread of epidemic disease on networks.
\newblock {\em Phys. Rev. E Stat. Nonlin. Soft Matter Phys.}, 66(1 Pt
  2):016128.

\bibitem[Newman, 2003]{Newman2003-is}
Newman, M. E.~J. (2003).
\newblock Mixing patterns in networks.
\newblock {\em Phys. Rev. E}, 67(2):026126.

\bibitem[Perkins et~al., 2018]{Tsai_Perkins2018-is}
Perkins, J.~M., Nyakato, V.~N., Kakuhikire, B., Tsai, A.~C., Subramanian,
  S.~V., Bangsberg, D.~R., and Christakis, N.~A. (2018).
\newblock Food insecurity, social networks and symptoms of depression among men
  and women in rural uganda: a cross-sectional, population-based study.
\newblock {\em Public Health Nutr.}, 21(5):838--848.

\bibitem[Perkins et~al., 2015]{Perkins2015-dm}
Perkins, J.~M., Subramanian, S.~V., and Christakis, N.~A. (2015).
\newblock Social networks and health: a systematic review of sociocentric
  network studies in low- and middle-income countries.
\newblock {\em Soc. Sci. Med.}, 125:60--78.

\bibitem[Randall et~al., 2013]{Randall2013-mb}
Randall, S.~M., Ferrante, A.~M., Boyd, J.~H., and Semmens, J.~B. (2013).
\newblock The effect of data cleaning on record linkage quality.
\newblock {\em BMC Med. Inform. Decis. Mak.}, 13:64.

\bibitem[Read et~al., 2008a]{Read2008-nh}
Read, J.~M., Eames, K. T.~D., and Edmunds, W.~J. (2008a).
\newblock Dynamic social networks and the implications for the spread of
  infectious disease.
\newblock {\em J. R. Soc. Interface}, 5(26):1001--1007.

\bibitem[Read et~al., 2008b]{Read2008-jl}
Read, J.~M., Eames, K. T.~D., and Edmunds, W.~J. (2008b).
\newblock Dynamic social networks and the implications for the spread of
  infectious disease.
\newblock {\em J. R. Soc. Interface}, 5(26):1001--1007.

\bibitem[Riley, 2007]{Riley2007-ur}
Riley, S. (2007).
\newblock Large-scale spatial-transmission models of infectious disease.
\newblock {\em Science}, 316(5829):1298--1301.

\bibitem[Rothenberg, 2009]{Rothenberg2009-el}
Rothenberg, R. (2009).
\newblock {HIV} transmission networks.
\newblock {\em Curr. Opin. HIV AIDS}, 4(4):260--265.

\bibitem[Salath{\'e} et~al., 2010]{Salathe2010-bp}
Salath{\'e}, M., Kazandjieva, M., Lee, J.~W., Levis, P., Feldman, M.~W., and
  Jones, J.~H. (2010).
\newblock A high-resolution human contact network for infectious disease
  transmission.
\newblock {\em Proc. Natl. Acad. Sci. U. S. A.}, 107(51):22020--22025.

\bibitem[Sariyar and Borg, 2010]{Sariyar2010-gi}
Sariyar, M. and Borg, A. (2010).
\newblock The {RecordLinkage} package: Detecting errors in data.
\newblock {\em R J.}, 2(2).

\bibitem[Shah et~al., 2017]{Shah2017-tj}
Shah, N.~S., Auld, S.~C., Brust, J. C.~M., Mathema, B., Ismail, N., Moodley,
  P., Mlisana, K., Allana, S., Campbell, A., Mthiyane, T., Morris, N.,
  Mpangase, P., van~der Meulen, H., Omar, S.~V., Brown, T.~S., Narechania, A.,
  Shaskina, E., Kapwata, T., Kreiswirth, B., and Gandhi, N.~R. (2017).
\newblock Transmission of extensively {Drug-Resistant} tuberculosis in south
  africa.
\newblock {\em N. Engl. J. Med.}, 376(3):243--253.

\bibitem[Takada et~al., 2019]{Takada2019-wr}
Takada, S., Nyakato, V., Nishi, A., O'Malley, A.~J., Kakuhikire, B., Perkins,
  J.~M., Bangsberg, D.~R., Christakis, N.~A., and Tsai, A.~C. (2019).
\newblock The social network context of {HIV} stigma: Population-based,
  sociocentric network study in rural uganda.
\newblock {\em Soc. Sci. Med.}, 233:229--236.

\bibitem[Thompson and Frank, 2000]{Thompson2000-fk}
Thompson, S.~K. and Frank, O. (2000).
\newblock Model-based estimation with link-tracing sampling designs.
\newblock {\em Surv. Methodol.}, 26(1):87--98.

\bibitem[Yamanis et~al., 2017]{Yamanis2017-kw}
Yamanis, T.~J., Dervisevic, E., Mulawa, M., Conserve, D.~F., Barrington, C.,
  Kajula, L.~J., and Maman, S. (2017).
\newblock Social network influence on {HIV} testing among urban men in
  tanzania.
\newblock {\em AIDS Behav.}, 21(4):1171--1182.

\end{thebibliography}

\bibliographystyle{apalike}

\end{document}